\documentclass[12pt]{article}
\usepackage{graphicx}

\begin{document}

\title{Space telescope \\ designed for accurate measurements}
\author{V.~Yu.~Terebizh\thanks{E-mail: valery@terebizh.ru}\\
Crimean Astrophysical Observatory, \\ Nauchny, Crimea 298409, Ukraine}

\date{}

\maketitle

\begin{abstract}
A modified version of the folded aplanatic Gregory telescope equipped with 
a spherical two-lens corrector is proposed for observations requiring a high 
signal-to-noise ratio. The basic telescope model has an aperture of 400~mm 
(f/3.0), its field of view is 3.$^\circ$0, the linear obscuration is 0.12, the 
distortion is less than 0.5\%. The focal surface has a spherical shape; the 
achieving of a plane field requires an increase in the number of lenses in 
the corrector. The images of stars in the integrated wavelength range 
0.35~- 1.0~$\mu$m are close to the diffraction-limited ones ($D_{80}$ = 
5.9~- 8.2~$\mu$m  = 1.$''$0~- 1.$''$4). The system is free from direct 
background illumination; both the lens corrector and the light detector are 
protected from cosmic particles. 
\end{abstract}

\section{Introduction} 

To solve some astronomical problems, one needs a telescope whose characteristics seem, 
at first glance, to be mutually incompatible: a significant field of view with image 
quality close to the diffraction limit; the small obscuration of useful light; the wide 
spectral range extending from the ultraviolet to the infrared region of the spectrum; 
the absence of background illumination; the simplicity of optical surfaces and the 
resulting comparative softness of tolerances on the parameters of the system. As an 
example of such problems, we point out the space project MESSIER (Valls-Gabaud 2016; 
Hugot et al. 2014), aimed at searching for a cosmological structure of extremely low 
surface brightness. This project further extends the requirements for the telescope by 
limiting the distortion of the image and refusing to use lenses because of the Cherenkov 
glow that appears in them when relativistic particles pass through. 

\begin{figure}   
   \centering
   \includegraphics[width=80mm]{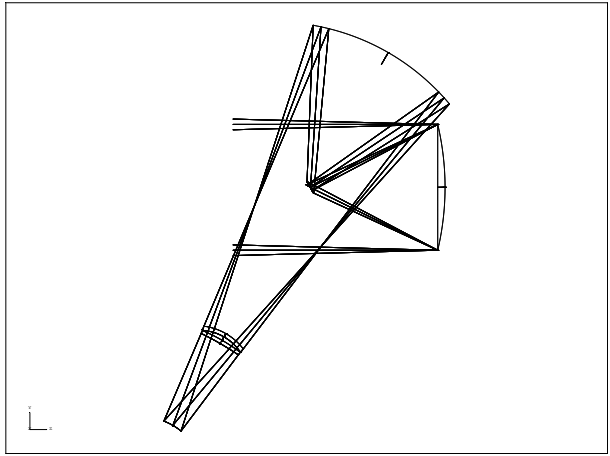}
   \caption{Optical layout of the VT-133c design.}
\end{figure}

Obviously, in these conditions it is not easy to find a suitable solution both 
in the class of axisymmetrical wide-field systems (Terebizh 2011, 2016) and 
among off-axial systems with aspheric and freeform optical surfaces (Hugot et al. 
2014; Buffington 1998; Singaravelu and Cabanac 2014; Challita et al. 2014).
In this note, we consider a 400-mm model, which is a modified version of the 
folded aplanatic Gregory telescope\footnote{
The aplanatic version of the Gregory system was proposed by Maksutov (1932). 
He also used the folding of the optical axis of the Gregory system in a 
telescope mounted in the 1940s at the Yerevan Observatory (Maksutov 1946).} 
provided by a two-lens corrector protected from cosmic particles. Although the 
model was designed to meet, as far as possible, the entire set of requirements 
listed above, we did not mean any specific project but wanted to give a general 
idea of the optical layout of the desired instrument.

\begin{table}
\caption{Basic characteristics of the VT-133c design}
\begin{tabular}{lc}
\hline \noalign{\smallskip}
    \textbf{Parameter}           & \textbf{Value} \\ 
\hline 
Entrance pupil diameter      & 400~mm  \\
Effective focal length       & 1200.2~mm  (5.82~$\mu$m/arcsec)  \\
Effective focal ratio        & 3.0     \\
Field of view diameter       & 3.$^\circ$0     \\
Primary spectral waveband    & 0.35 - 1.0~$\mu$m   \\
Linear obscuration           & 0.120    \\	
Fraction of unvignetted light, center - edge  & 0.984 - 0.984 \\
Effective aperture diameter  &	397~mm \\
Maximum distortion           &	0.48\% \\	
Spot diameter in integral light, center - edge    & 
     2.8 - 5.1~$\mu$m  (0.$''$48 - 0.$''$88)) \\
Diameter of a circle containing 80\% of energy in  &    \\ 
a star image (D$_{80}$). Integral light, center - edge & 
     5.9 - 8.2~$\mu$m  (1.$''$0 - 1.$''$4) \\
Maximum lens diameter                    &	145~mm \\
Curvature radius of the image surface    &  195~mm  \\
\noalign{\smallskip}\hline
\end{tabular}
\end{table}

\section{The design}  

The model VT-133c is shown in Figure~1 and is described in Tables~1 and~2. 
Optical surfaces of power mirrors are ellipsoids; the primary mirror is close 
to the paraboloid, and the secondary mirror~-- to the sphere. Two spherical 
lenses of the corrector are made of fused silica. 

The insignificant obscuration of the incoming light flux (1.6\%) and the absence 
of direct background illumination are provided by folding of the optical layout 
with the aid of a small flat mirror. This same feature maintains the axial 
symmetry of the system, which significantly simplifies the optical surfaces and 
makes the tolerances on the parameters far less tight in comparison with off-axis 
systems. In addition, the folding of layout makes it possible to reliably shield 
the lenses and the light detector from cosmic particles. To fasten a fold mirror 
(in this example, of a diameter less than 43~mm), thin stretches can be used that 
introduce negligible light diffraction. 

\begin{figure}      
\includegraphics[width=130mm]{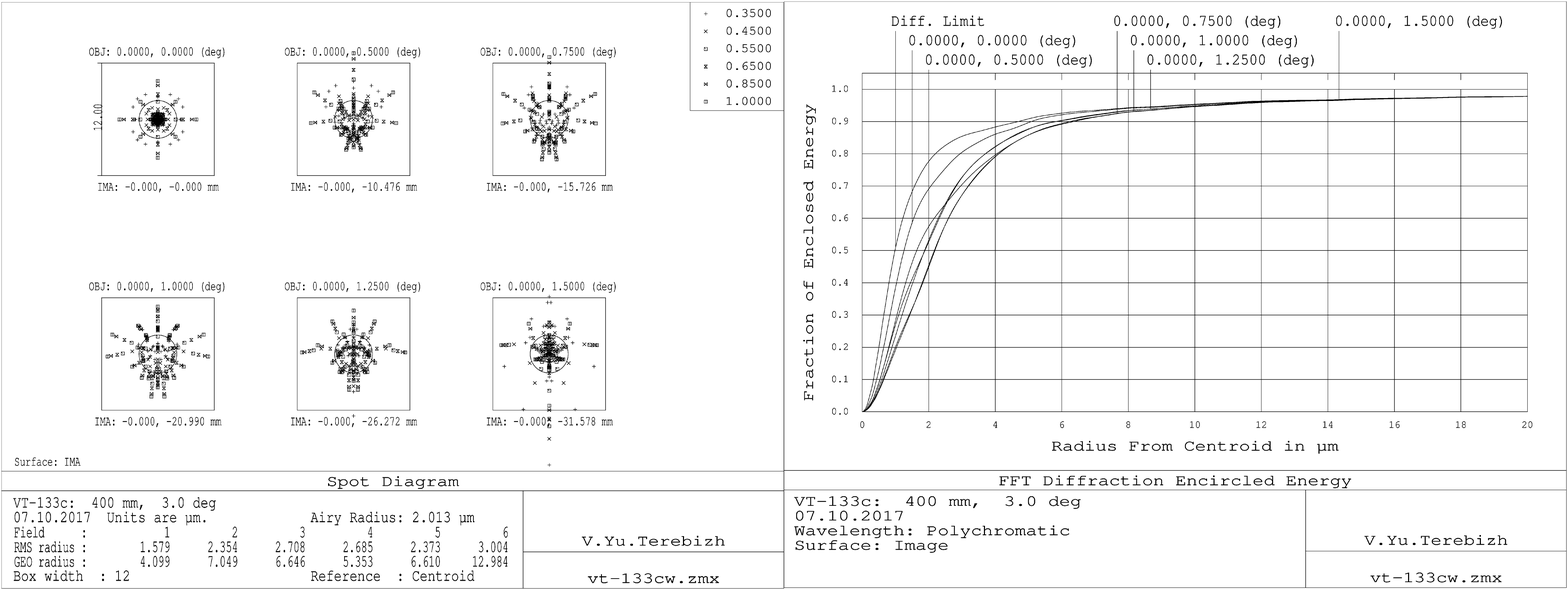} 
	\caption{Spot diagrams (on the left) and diffraction encircled energy 
		distributions (on the right) for the VT-133c design in polychromatic 
		light 0.35~- 1.0~$\mu$m. Width of the boxes on the spot diagrams is 
		12~$\mu$m. The Airy diffraction disc is shown by a circle.} 
\end{figure}

Table~2 gives a complete description of the optical scheme of VT-133c in the format 
adopted by {\it Zemax} optics calculation program. The following notation is used: 
$R_0$~-- the paraxial curvature radius; $T$~-- the distance to the next surface; 
$D$~-- the current light aperture; Stop~-- the aperture stop; Pri~-- the primary 
mirror; Sec~-- the secondary mirror; CB~-- the coordinate break; $L_1$, $L_2$~-- 
lenses No.~1 and No.~2; Ima~-- the focal surface; FS~-- fused silica. 

As shown in Fig.~2, the spot diameter of a star image in the polychromatic light 
0.35 - 1.0~$\mu$m does not exceed 5.1~$\mu$m at the image scale of 
5.8~$\mu$m/arcsec and the diameter of the Airy disc of 4.9~$\mu$m for the central 
wavelength. The $D_{80}$ diameter of images is less than 8.2~$\mu$m along the field 
of view. Thus, the system under consideration provides nearly diffraction-limited 
images in a sufficiently wide field of view and an extended spectral range. 

The fold mirror obscures both the incoming light beam and the beam reflected by 
the secondary mirror; the latter factor is more significant. Angle of fracture 
of the initial optical axis, $-30^\circ$, is chosen approximately, so a 
small reduction in obscuration can be achieved by optimizing this parameter. 

The choice of fused silica as a material for lenses is only partly due to the 
excellent optical properties of this glass; many other types of glass are also 
suitable. 

We did not touch on the secondary issues involved in choosing the appropriate 
thickness of lenses and flat filters. 

It is also worth noting that the cross-shaped optical scheme of the telescope 
makes it possible to realize a fairly compact design.

\begin{table} 
\caption{VT-133c design with 400~mm aperture and 3.$^\circ$0 field of view.}
\begin{tabular}{cccccc}
\hline \noalign{\smallskip} 
Surf.& Com-   & $R_0$       & $T$         & Glass   & $D$    \\
 No. & ments  & (mm)        & (mm)        &         & (mm)   \\
\hline 
	1   & Stop       & $\infty$    & 23.259      & --      & 400.0 \\
	2   & Pri        & $-$862.191  & $-$427.829  & Mirror  & 400.0 \\
	3   & CB         &             &  0          & --      &  \\
	4   & Fold       & $\infty$    &  0          & Mirror  &  42.79 \\
	5   & CB         &             &  494.734    & --      &  \\
	6   & Sec        & $-$723.065  & $-$1031.839 & Mirror  & 500  \\ 	
	7   & L$_1$      & $-$187.284  & $-$12.0     & FS      & 144.4  \\
	8   &            & $-$156.714  & $-$12.338   & --      & 138.6  \\
	9   & L$_2$      & $-$559.139  & $-$12.0     & FS      & 138.5  \\
	10  &            & $-$996.772  & $-$300.545  & --      & 136.4  \\
	11  & Ima        & $-$194.795  & --          & --      &  63.2  \\
\noalign{\smallskip}\hline
\end{tabular} 

Notes to Table~2:\\
Conic constants of the primary and secondary mirrors are, respectively,
$-$0.802053 and $-$0.292905; all other surfaces are spheres. 
Tilt about X-axis for the surfaces No.~3 and No.~5 is $-30^\circ$. 
\end{table}

\section{Conclusions} 

Evidently, replacing the two-lens corrector with a concave mirror in the system 
under discussion will result in a three-mirror telescope of the type described 
long ago by Dimitroff and Baker (1945), although with dissimilar types of mirror 
surfaces (see also Wilson 1996). In this way, it is possible to provide high 
quality images on a spherical focal surface, but the linear obscuration of light 
in the system rises at least to 0.25. The latter is not only undesirable in 
itself, but also worsens the concentration of energy in a star's image. 

The most obliging features of the proposed system are the curvature of 
the focal surface, the large size of the secondary mirror, and the significant 
asphericity of both mirrors with optical power. 

The first point cannot now be regarded as a serious drawback of the optical layout. 
In the last decade, many exploratory and working detectors with a spherical surface 
have been manufactured. Since this issue was discussed in a number of publications 
(see, e.g., Rim et al. 2008; Dinyari et al. 2008; Lesser and Tyson 2002; Iwert et 
al. 2012, and the references in Terebizh 2016), we will not go into details here. 
On the other hand, it is possible to achieve a flat focal surface by increasing 
the number of lenses in the corrector to four, using suitable glass grades for them 
(say, from a recommended Ohara's list) and slightly relaxing the restriction on 
the amount of distortion. As a result, the system becomes more complicated, so 
it is reasonable to ask about general priorities. 

As for the second of the problems mentioned above, the significant size of the 
secondary mirror seems to be an unavoidable feature of this layout at low $f$-number  
and a large field of view. Only because of this a two-mirror, in essence, system 
attains such a low obscuration of useful light in the absence of direct 
background illumination.

Finally, let us turn to the third point. The main problem that arises in making 
of axisymmetric aspheric surfaces is not related to the magnitude of the deviation 
from a nearest sphere but rather to the {\it asphericity gradient} $G$ ($\mu$m/mm), 
i.e., the rate at which the deviation changes along the radial coordinate. An 
approximate expression for the maximum asphericity gradient $G_{max}$ of a conic 
section as a function of its diameter $D$, the radius of curvature at the vertex 
$R_0$ and the eccentricity $\varepsilon$ is: 
$$
  G_{max} \simeq 31.25\, \varepsilon^2 (D/|R_0|)^3,\quad \mu m/mm
$$
(Terebizh 2011). For mirrors, it is simpler to use the equivalent formula: 
$$
  G_{max} \simeq 3.906\, |k|/\phi^3,\quad \mu m/mm,
$$
where $k \equiv -\varepsilon^2$ is a conic constant, and $\phi \equiv R_0/(2D)$ 
is a focal ratio. In our case, according to Table~2, the maximum asphericity 
gradient of the primary mirror is 2.5~$\mu$m/mm, and of the secondary mirror~-- 
3.0~$\mu$m/mm. These are large values, but they are within the limits of modern 
technology capabilities. For example, the primary and secondary mirrors of the 
1.8-meter Vatican Advanced Technology Telescope (West et al. 1997), 
which was put into operation in 1993, have maximum asphericity gradient 
3.9~$\mu$m/mm and 3.6~$\mu$m/mm, respectively. These mirrors were successfully 
manufactured by the University of Arizona's Steward Observatory Mirror Laboratory 
and the Space Optics Research Laboratory (Chelmsford, MA).

It can be hoped that the optical layout discussed here will be useful not only 
for space telescopes but also for ground-based ones. 

\subsection*{Acknowledgments}

I thank M.R.~Ackermann (University of New Mexico, USA), Yu.A.~Petrunin (Telescope 
Engineering Company, USA) and V.N.~Skiruta (Crimean Astrophysical Observatory, 
Ukraine) for useful comments on the note.

\end{document}